\documentclass[useAMS,natbib,multicolumn,subfigure]{mn2e}
\usepackage{amsmath}
\usepackage[T1]{fontenc}
\usepackage{ae,aecompl}

\usepackage[latin1]{inputenc}  
\usepackage{graphicx,amsmath,xcolor}
\usepackage{amssymb}
\usepackage{indentfirst}
\usepackage{hyperref}
\usepackage{graphicx,pst-all}
\usepackage{tikz}
\usepackage{layout}
\usepackage{fancyhdr}

\newcommand{\kms}{\hbox{\kern 0.20em km\kern 0.20em s$^{-1}$}}
\newcommand{\cmt}{\hbox{\kern 0.20em cm$^{-3}$}}
\newcommand{\cmd}{\hbox{\kern 0.20em cm$^{-2}$}}

\def\msol{\hbox{\kern 0.20em $M_\odot$}}
\def\lsol{\hbox{\kern 0.20em $L_\odot$}}
\def\rsol{\hbox{\kern 0.20em $R_\odot$}}
\def\sr{\hbox{\kern 0.20em sr}}
\def\srmu{\hbox{\kern 0.20em sr$^{-1}$}}

\def\g{\hbox{\kern 0.20em g}}
\def\gmu{\hbox{\kern 0.20em g$^{-1}$}}
\def\kg{\hbox{\kern 0.20em kg}}
\def\pc{\hbox{\kern 0.20em pc}}

\def\mum{\hbox{\kern 0.20em $\mu$m}}
\def\mumd{\hbox{\kern 0.20em $\mu$m$^{-2}$}}
\def\cm{\hbox{\kern 0.20em cm}}
\def\m{\hbox{\kern 0.20em m}}
\def\km{\hbox{\kern 0.20em km}}
\def\nm{\hbox{\kern 0.20em nm}}

\def\s{\hbox{\kern 0.20em s}}
\def\h{\hbox{\kern 0.20em h}}
\def\sec{\hbox{\kern 0.20em sec}}
\def\min{\hbox {\kern 0.20em min}}
\def\smu{\hbox{\kern 0.20em s$^{-1}$}}
\def\smd{\hbox{\kern 0.20em s$^{-2}$}}
\def\an{\hbox{\kern 0.20em an}}
\def\anmu{\hbox{\kern 0.20em an$^{-1}$}}
\def\deg{\hbox{\kern 0.20em $^{\rm o}$}}
\def\yr{\hbox{\kern 0.20em yr}}
\def\yrmu{\hbox{\kern 0.20em yr$^{-1}$}}
\def\Myr{\hbox{\kern 0.20em Myr}}
\def\Mymu{\hbox{\kern 0.20em Myr$^{-1}$}}
\def\K{\hbox{\kern 0.20em K}}
\def\pcmu{\hbox{\kern 0.20em pc$^{-1}$}}
\def\pcmd{\hbox{\kern 0.20em pc$^{-2}$}}
\def\pcmt{\hbox{\kern 0.20em pc$^{-3}$}}
\def\kms{\hbox{\kern 0.20em km\kern 0.20em s$^{-1}$}}
\def\kmpd{\hbox{\kern 0.20em km$^{2}$}}
\def\kpc{\hbox{\kern 0.20em kpc}}
\def\cms{\hbox{\kern 0.20em cm\kern 0.20em s$^{-1}$}}
\def\erg{\hbox{\kern 0.20em erg}}
\def\ergs{\hbox{\kern 0.20em erg}}
\def\cmpd{\hbox{\kern 0.20em cm$^2$}}
\def\cmmd{\hbox{\kern 0.20em cm$^{-2}$}}
\def\cmms{\hbox{\kern 0.20em cm$^{-6}$}}
\def\cmpt{\hbox{\kern 0.20em cm$^3$}}
\def\cmmt{\hbox{\kern 0.20em cm$^{-3}$}}
\def\mpd{\hbox{\kern 0.20em m$^2$}}
\def\mmd{\hbox{\kern 0.20em m$^{-2}$}}
\def\mpt{\hbox{\kern 0.20em m$^3$}}
\def\mmt{\hbox{\kern 0.20em m$^{-3}$}}
\def\mujy{\hbox{\kern 0.20em $\mu$Jy}}
\def\mjy{\hbox{\kern 0.20em mJy}}
\def\Mj{\hbox{\kern 0.20em MJy}}
\def\jy{\hbox{\kern 0.20em Jy}}
\def\ghz{\hbox{\kern 0.20em GHz}}
\def\srmd{\hbox{\kern 0.20em sr$^{-1}$}}

\def \mum{$\mu$m}
\def\G{\hbox{\kern 0.20em G}}

\def\htwo{\hbox{H${}_2$}}
\def\h13cop{\hbox{H$^{13}$CO$^{+}$}}

\def\h2o{\hbox{H$_2$O}}

\begin{document}

\title[]{L1157-B1, a factory of complex organic molecules in a Solar-type
star forming region}

\author[B. Lefloch et al.]{
Bertrand Lefloch$^{1}$\thanks{E-mail: bertrand.lefloch@univ-grenoble-alpes.fr},
C. Ceccarelli$^{1}$,
C. Codella$^{2}$,
C. Favre$^{2,1}$,
L. Podio$^{2}$,
\newauthor
C. Vastel$^{3}$,
S. Viti$^{4}$,
R. Bachiller$^{5}$
\\
$^1$Univ. Grenoble Alpes, CNRS, IPAG, 38000 Grenoble, France\\
$^2$INAF, Osservatorio Astrofisico di Arcetri, Largo Enrico Fermi 5, I-50125 Firenze, Italy\\
$^3$Universit\'e de Toulouse, UPS-OMP, IRAP, Toulouse, France\\
$^4$Department of Physics and Astronomy, UCL, Gower St., London, WC1E 6BT, UK\\
$^5$IGN Observatorio Astronómico Nacional, Apartado 1143, 28800 Alcalá de Henares, Spain\\
}

\date{Accepted 2017 April 3. Received 2017 April 1; in original form 2017 February 18}
\pubyear{2017}

\label{firstpage}
\pagerange{\pageref{firstpage}--\pageref{lastpage}}
\maketitle

\begin{abstract}
  We report on a systematic search for oxygen-bearing Complex Organic
  Molecules (COMs) in the Solar-like protostellar shock region
  L1157-B1, as part of the IRAM Large Program "Astrochemical Surveys At IRAM" (ASAI). Several COMs are
  unambiguously detected, some for the first time, such as ketene H$_2$CCO, dimethyl
  ether (CH$_3$OCH$_3$) and glycolaldehyde (HCOCH$_2$OH), and others
  firmly confirmed, such as formic acid (HCOOH) and ethanol
  (C$_2$H$_5$OH). Thanks to the high sensitivity of the observations
  and full coverage of the 1, 2 and 3mm wavelength bands, we detected
  numerous ($\sim$10--125) lines from each of the detected
  species. Based on a simple rotational diagram analysis, we derive
  the excitation conditions and the column densities of the detected
  COMs. Combining our new results with those previously obtained
  towards other protostellar objects, we found a good correlation between
  ethanol, methanol and
  glycolaldehyde. We discuss the implications of these results on the possible
  formation routes of ethanol and glycolaldehyde.
\end{abstract}


\maketitle
\section{Introduction}

There is now substantial observational evidence that protostellar shocks can significantly alter the chemical composition of  the medium in which they propagate. Thanks to its remarkably bright molecular line emission, the region L1157-B1 (d=250 pc; Looney et al. 2007) has been the privileged target of many observational studies to investigate the physical and chemical properties of protostellar shocks. Gueth et al. (1996) showed that the L1157-B1 shock was caused by the high-velocity, precessing jet driven by the low luminosity Class 0 source L1157-mm (L$\sim 4\lsol$). Following the pioneering study of Bachiller \& Perez-Gutierrez (1997), Arce et al. (2008) reported the detection of a few COMs: methyl formate (HCOOCH$_3$), methyl cyanide (CH$_3$CN), ethanol (C$_2$H$_5$OH), and formic acid (HCOOH).

Since then, individual studies have reported the presence of various organic species in that region (e.g. Codella et al. 2010). Codella et al. (2009) proved unambiguously  that CH$_3$CN was associated with the shock itself. More recently,  Mendoza et al. (2014) have reported the detection of formamide NH$_2$CHO, with an abundance ranging among the highest values ever reported in molecular clouds (Bisschop et al. 2007; Lopez-Sepulcre et al. 2015). This detection raises the question of the molecular complexity which is ultimately reached in protostellar shocks and the nature of the involved chemical processes, either in the gas phase or at the surface of dust grains.

L1157-B1 was observed as part of ASAI\footnote{http://www.oan.es/asai/} with the IRAM 30m telescope  to investigate the feedback of shocks on the molecular gas.  In this Letter, we present and discuss the results of our search for Oxygen-bearing COMs. From a highly sensitive, unbiased spectral line survey of the 72--350 GHz band, we have obtained a comprehensive census of the COMs content in L1157-B1. The detected lines span a wide range of excitation, with $E_{up}$ ranging from about $5\K$ up to 40--180\K,
and spontaneous emission coefficients $A_{ij}$,  from $10^{-6}\smu$ to a few $10^{-4}\smu$, depending on the species.
This has allowed us to determine with unprecedented accuracy the excitation conditions and abundances of these species  in L1157-B1.
%

\section{Observations}
\begin{table*}
\centering{
\caption{Oxygen-bearing Complex Organic Molecules detected  with ASAI towards the
    protostellar shock~L1157-B1: methanol (CH$_3$OH), ketene(H$_2$CCO), ethanol(C$_2$H$_5$OH), acetaldehyde (CH$_3$CHO), methyl formate (HCOOCH$_3$), glycolaldehyde (HCOCH$_2$OH), dimethyl ether (CH$_3$OCH$_3$), formic acid (HCOOH), and formamide (NH$_2$CHO).  For each molecular species, we indicate the database and the TAG used, the range of $E_{up}$ and $A_{ij}$ spanned by the transitions detected, the number of detected lines, the excitation temperature and the column density obtained from a
    simple rotational diagram analysis, and the derived abundance with respect to $\htwo$. We adopt the convention $a(b)= a\times 10^b$.}
  \label{tab:obs-results}
\begin{tabular}{lrrrccrrr}
\hline
Species             & Database & TAG & Lines & $E_{up} (\K)$ &  $A_{ij}(\smu)$ &  $T_{rot}(\K) $   & N ($\cmmd$)    &   X    \\
    \hline
CH$_3$OH      & JPL & 32003 &125& 4.6 -- 177.5 & 2.0(-6) -- 2.3(-4)& $25.1\pm 1.5$ & ($10\pm 1.0$)(14)& ($5.0\pm 1.0$)(-7) \\
H$_2$CCO      & CDMS& 42501 & 23& 9.7 -- 105.5 & 5.0(-6) -- 2.1(-4)& $23.0\pm 1.1 $&($1.7\pm 0.7$)(13)& ($8.5\pm 4.3$)(-9)\\
C$_2$H$_5$OH  & JPL & 46004 & 22& 9.3 -- 74.4  & 5.1(-6) -- 8.5(-5)& $28.2\pm 4.5$& ($4.6\pm 1.1$)(13)& ($2.3\pm 0.8$)(-8) \\
CH$_3$CHO     & JPL & 44003 & 78& 11.8 -- 94.1 & 1.5(-5) -- 5.2(-4)& $16.6\pm 0.7$& ($2.8\pm0.4$)(13) & ($1.4\pm 0.4$)(-8)\\
HCOOCH$_3$    & JPL & 60003 & 57& 20.1 -- 77.7 & 1.0(-5) -- 6.5(-5)& $20.7\pm 1.5$& ($5.4\pm 0.8$)(13)& ($2.7\pm 0.7$)(-8)\\
HCOCH$_2$OH   & JPL & 60006 & 14& 18.8 -- 93.8 & 4.0(-6) -- 2.1(-5)& $50.4\pm10.2$ &($3.4\pm 0.7$)(13)& ($1.7\pm 0.5$)(-8) \\
HCOCH$_2$OH$^a$ & JPL & 60006 &  9& 18.8 -- 53.2 & 4.8(-6) -- 2.1(-5)& $31.1\pm 5.3$ &($2.1\pm 0.3$)(13)& ($1.0\pm 0.2$)(-8) \\
CH$_3$OCH$_3$ & JPL & 46008 & 31&  6.7 -- 40.4 & 1.7(-6) -- 2.1(-5)& $13.9\pm 1.6$ &($4.9\pm 0.9$)(13)& ($2.5\pm 0.7$)(-8)  \\
t-HCOOH       & JPL & 46005 & 38& 10.8 -- 110.2& 5.5(-6) -- 1.9(-4)& $21.2\pm 1.1$& ($1.7\pm0.2$)(13) & ($8.5\pm 1.9$)(-9) \\
NH$_2$CHO$^b$ & CDMS& 45512 & 16& 10.2 -- 32.5 & 9.2(-6) -- 2.5(-4)& $10.0\pm 2.0$&($3.5\pm 1.0$)(12) & ($1.8\pm 0.7$)(-9)  \\
              & CDMS& 45512 &  8& 60.8 -- 82.9 & 2.1(-5) -- 1.0(-3)& $34.0\pm 2.0$&($1.7\pm0.2$)(12)  & ($8.5\pm 1.9$)(-10) \\
\hline
\end{tabular}
}
\\ $^a$ Only lines with $E_{up} \leq 55\K$ are considered. $^b$ Two excitation components are detected (Mendoza et al. 2014).\\
\end{table*}

The observations of L1157-B1 were carried out with the IRAM-30m
telescope at Pico Veleta (Spain), during several runs in 2011 and
2016, and covered the spectral bands 72 -- 116 GHz, 128 -- 173 GHz,
200 -- 260 GHz, and 260 -- 350 GHz.  The observed position of L1157-B1 is
$\alpha_{J 2000}=$ 20$^{\text h}$39$^{\text m}$10.$^{\text s}$2,
$\delta_{J 2000} =$ +68$^{\circ}$01$^{\prime}$10$^{\prime\prime}$.
The survey was carried out using the broad-band
EMIR receivers. Fast Fourier Transform Spectrometers were connected to
the EMIR receivers, providing a spectral resolution of 195 kHz. The
high-frequency part  band 260--350 GHz was observed with
the WILMA autocorrelator, at 2 MHz resolution. The final kinematic
resolution of the FTS data was degraded to $1\kms$. A description of
the observing procedure and data acquisition  will be presented more
extensively in the article
presenting the ASAI survey (Lefloch in prep).

The data reduction was performed using the GILDAS/CLASS90
package\footnote{http://www.iram.fr/IRAMFR/GILDAS/}. The line
intensities are expressed in units of antenna temperature corrected
for atmospheric attenuation and rearward losses ($T_A^{\star}$). For
the ASAI data, the calibration uncertainties are typically 10, 15, and
$20\%$ at 3mm, 2mm, and 1.3mm, respectively (Lefloch in prep). For subsequent analysis,
fluxes were expressed in main beam temperature units ($T_{mb}$). The
telescope and receiver parameters (main-beam efficiency Beff, forward
efficiency Feff, Half Power beam Width HPBW) were taken from the IRAM
webpage\footnote{http://www.iram.es/IRAMES/mainWiki/Iram30mEfficiencies}. The
rms noise achieved is typically 1mK, 3--5mK and 2--5mK ($T_{A}^{*}$)
in a velocity interval of $1\kms$ range at 3mm, 2mm, and 1.3mm,
respectively.

\section{Results and discussion}

\begin{figure*}
  \begin{center}
    \caption[]{Montage of detected transitions from the different O-bearing COMs detected towards L1157-B1. Intensities are
      expressed in units of $T_A^{*}$. The red dashed line marks the
      ambient cloud velocity $v_{lsr}=+2.6\kms$. Predicted line intensities in the LTE regime and based on our
      rotational diagram analysis are shown in red.}
    \label{fig:spectra}
    \includegraphics[width=1.5\columnwidth]{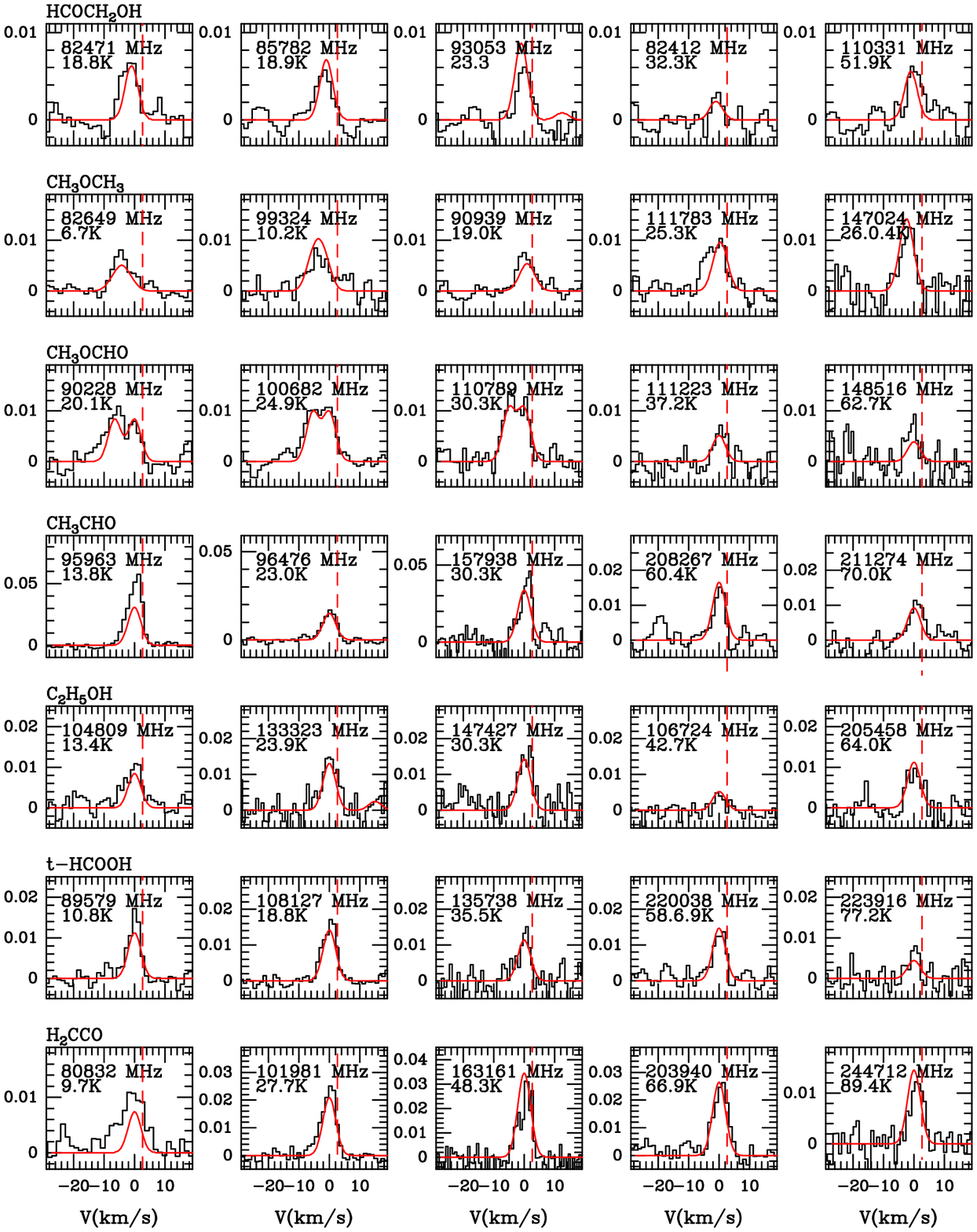}
  \end{center}
\end{figure*}

\subsection{COMs in L1157-B1}
We performed a systematic search from the list of Oxygen-bearing COMs identified in
the sun-like protostar IRAS16293-2422 (Jaber et al 2014). We used the
CASSIS software (Vastel et al. 2015) for the line
identification, using the CDMS and JPL databases.
We detected nine Oxygen-bearing COMs: methanol (CH$_3$OH), ketene (H$_2$CCO), ethanol (C$_2$H$_5$OH), acetaldehyde (CH$_3$CHO),
methyl formate (HCOOCH$_3$), glycolaldehyde (HCOCH$_2$OH), dimethyl ether (CH$_3$OCH$_3$),
formic acid (HCOOH), and formamide (NH$_2$CHO). We also detected the rare isotopologues ($^{13}$C, $^{18}$O)
of methanol. Dimethyl ether, glycolaldehyde and ketene are detected for the first
time in a protostellar shock. In addition, two other species,
whose identification was based on one or two lines (Arce et
al. 2008), are now firmly detected: ethanol  and formic acid. The detection of formamide was previously reported and discussed by Mendoza et al. (2014).  Our results are summarized in Table \ref{tab:obs-results}.

For each detected species, we could identify from 14
transitions (glycolaldehyde) up to more than 100 transitions
(methanol). A montage of illustrative transitions from each species (but methanol) is displayed in
Fig.~1.  Line intensities are weak, typically ranging
between 10 and 20 mK ($\rm T_{A}^{*}$), with the exception of
acetaldehyde, whose transitions display higher $A_{ij}$, and methanol. All the lines are blueshifted with respect to the cloud
ambient velocity $v_{lsr}$= $+2.6\kms$ (the peak intensity lies at $+0\kms$)
and display typical linewidths
of $\sim 4$ to $7\kms$ (FWHM). This supports evidence for emission from the
shocked material of L1157-B1 and not from the ambient molecular
cloud. Recently, using the PdBI at $2\farcs5$ resolution, Codella et
al. (2015) showed that the emission of CH$_3$CHO  is associated with the outflow cavity of L1157-B1.
The similarity of the line profiles (see Fig.~1)
suggests that this is probably the case for all the COMs
reported here. These detections are all the more remarkable as we did not find any evidence of COM
emission towards L1157-mm itself, the protostar at the origin of the outflow phenomenon and its
associated shocks (Tafalla et al. 2015; Podio et al. 2016)



\subsection{Molecular abundances}
\begin{figure}
  \begin{center}
    \includegraphics[width=0.7\columnwidth]{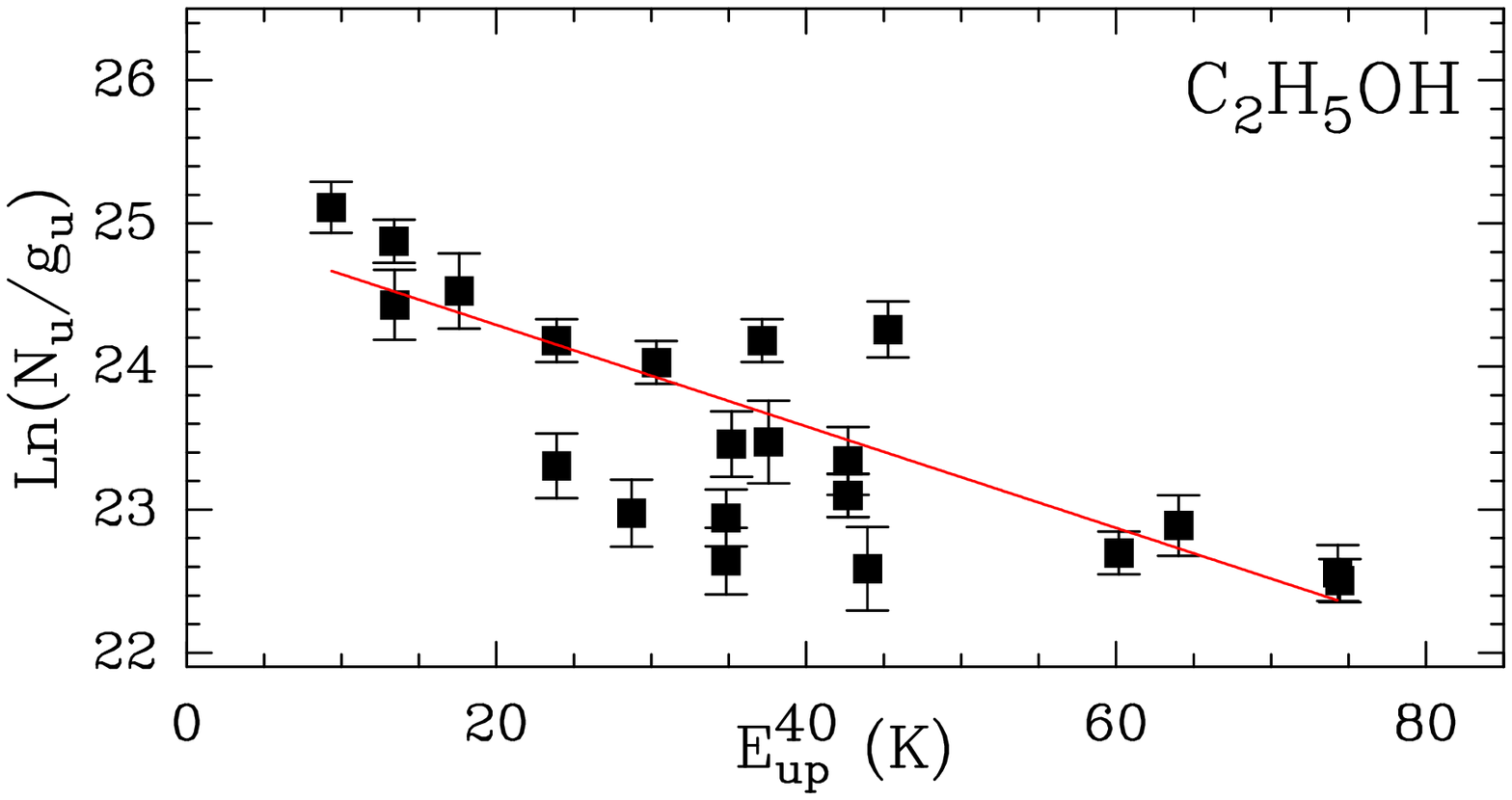}
    \includegraphics[width=0.7\columnwidth]{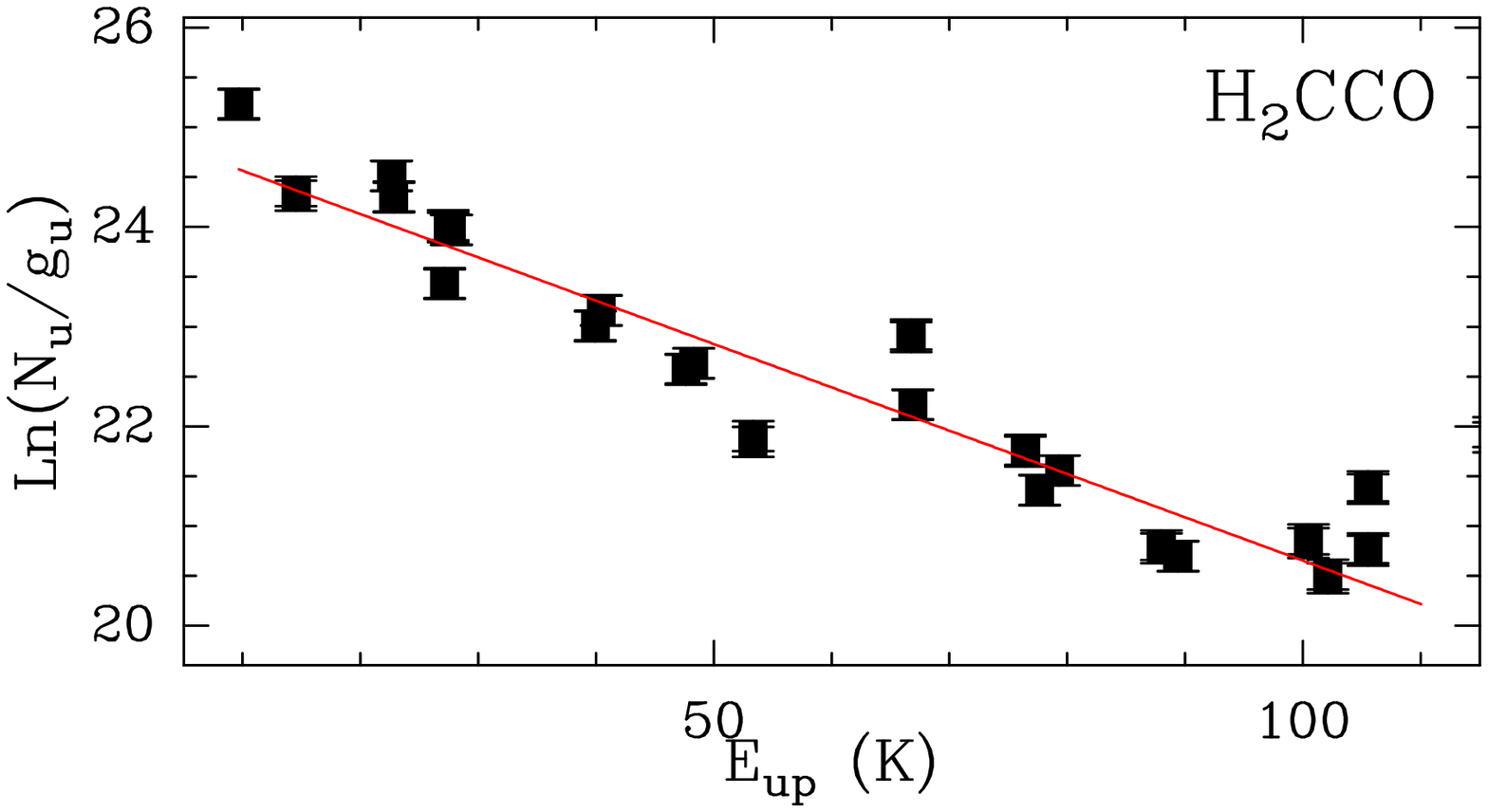}
    \includegraphics[width=0.7\columnwidth]{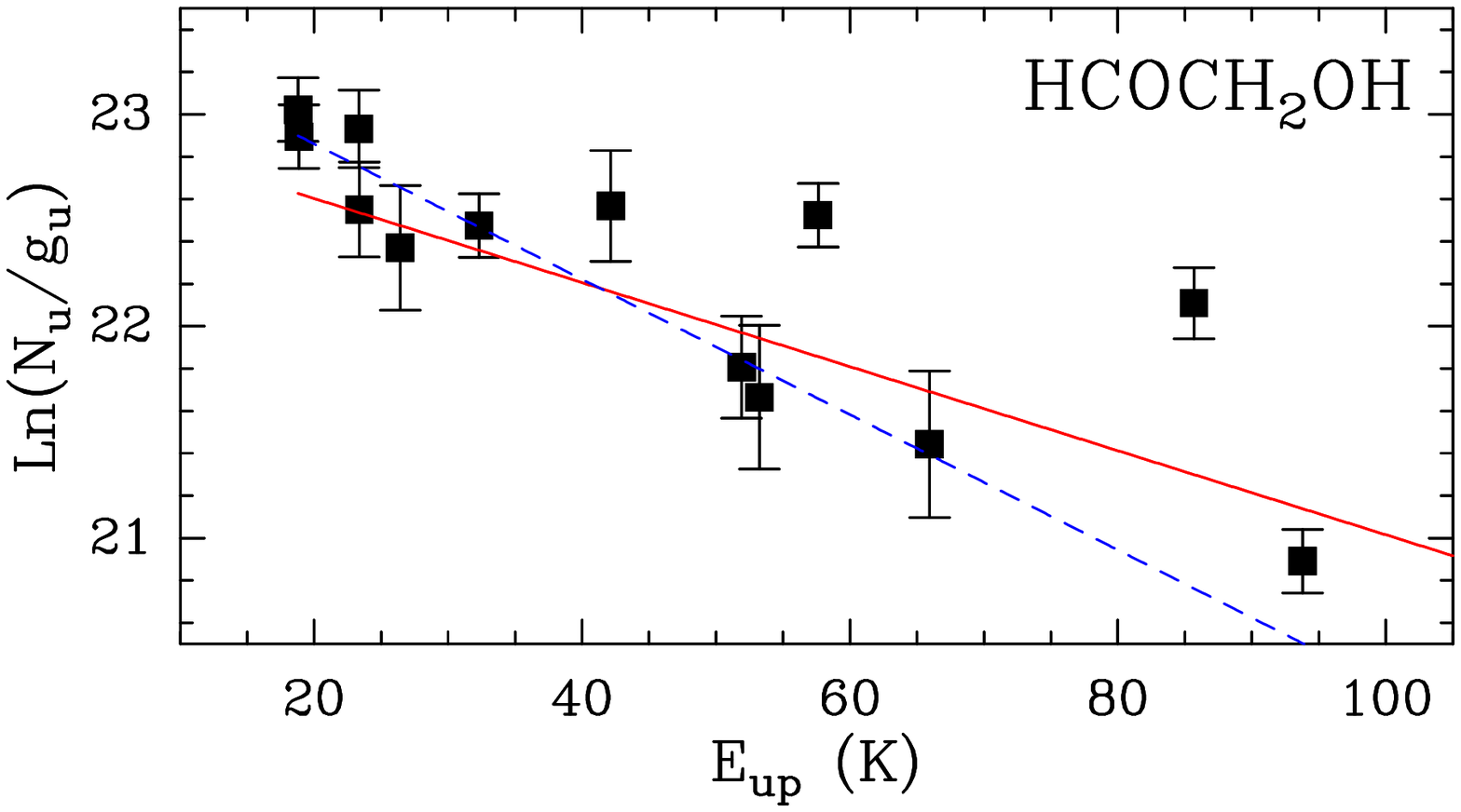}
    \includegraphics[width=0.7\columnwidth]{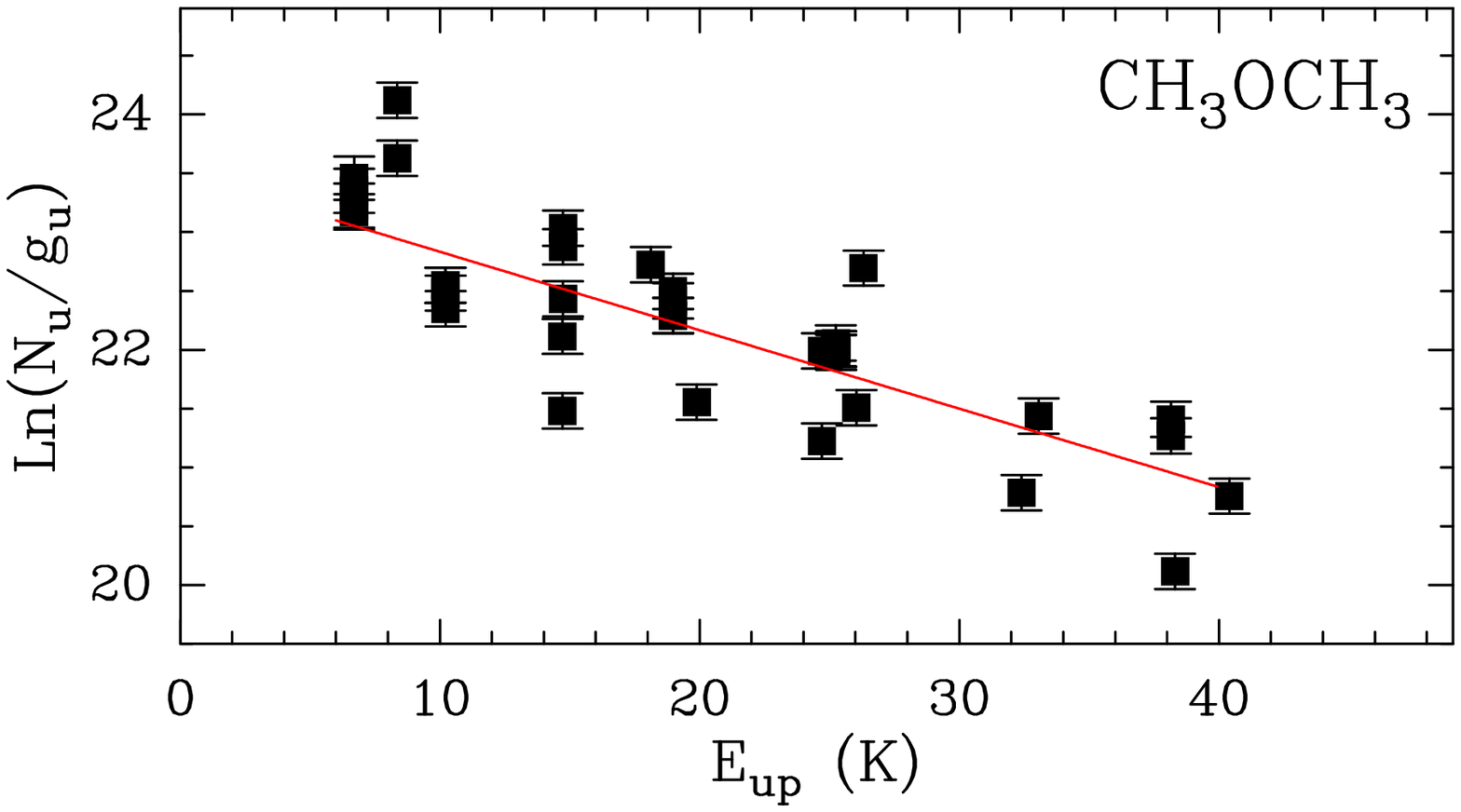}
    \caption[]{Rotational diagrams of COMs detected towards L1157-B1. From top to bottom: C$_2$H$_5$OH, H$_2$CCO, HCOCH$_2$OH and CH$_3$OCH$_3$. The best fit is drawn by the red line. For HCOCH$_2$OH, the fit to the transitions with  $E_{up}\leq 55\K$ is drawn by the blue dashed line.}
    \label{fig:rotdiag}
  \end{center}
\end{figure}

Given the large number ($\geq 10$) of detected lines for each species,
our new observations allow us not only to firmly establish the
presence but also to give reliable estimates of the relative
abundances of the COMs detected in L1157-B1. With this goal in mind, we
obtained their excitation conditions and column density carrying out a
simple rotational diagram analysis. We computed the molecules abundances with respect
to \htwo\ by using the {\em whole} CO column density in the B1 direction, as estimated by
Lefloch et al. (2012) N(CO)= $2\times 10^{17}\cmmd$ and assuming a standard relative CO abundance
X(CO)=N(CO)/N($\htwo$)= $10^{-4}$.
The obtained temperatures, column densities, and abundances are summarized in Table 1

The rotational diagrams of the newly
detected COMs
are shown in Fig.~\ref{fig:rotdiag}.
For all the detected species, the line emission can be reasonably well
fit by one single gas component whose rotational temperature is around
20--30 K (Tab. \ref{tab:obs-results}). Glycolaldehyde has a slightly
larger rotational temperature ($\sim50$ K) but also a larger error bar
($\sim 10$ K). Considering only the lines observed with Eup $\leq 55\K$, a
lower rotational temperature ($31.1\pm 5.3\K$) would provide a very good fit.

Column densities vary from $\sim 2\times 10^{12}$ cm$^{-2}$
(formamide) to $\sim 1\times 10^{15}$ cm$^{-2}$
(methanol). Incidentally, the column density of the other COMs shows a
relatively small range of variation, from $\sim2$ to $\sim5$
$\times 10^{13}$ cm$^{-2}$, namely less than a factor of three. In other
words, with the exception of formamide, which is significantly less
abundant, the other detected COMs have abundances a factor 20 to 50
lower than methanol.

We searched for acetic acid (CH$_3$COOH),
the other isomer of glycolaldehyde and methyl formate, using the frequencies given in Ilyushin et al. (2008). We failed to detect it. We could place an upper limit on its abundance, adopting $T_{rot}$ similar to that of methyl formate ($20\K$) and glycolaldehyde ($50\K$),  and the partition function
given by Calcutt et al. (in prep).  We obtain a $3\sigma$ upper limit of $5\times 10^{-9}$ at $20\K$ ($10^{-8}$ at $50\K$): acetic acid is less abundant by a factor of 3 to 5 with respect to methyl formate.

\subsection{Comparison with hot corinos}
When compared to other Solar-like star forming regions where these COMs were previously detected
(IRAS16293-2422: Jaber et al. 2014, Coutens et al. 2015, J{\o}rgensen et al. 2016; IRAS2A and IRAS4A: Taquet et al. 2015),
L1157-B1 appears to be a site were COMs are very efficiently produced.
The molecular abundances are higher in the shock than in the  hot corinos  by a factor of 2 to 10, depending on the species. The dimethyl ether abundance measured in IRAS16293-2422 is the only exception. We note that only relatively small COMs are detected in the shock region, whereas larger COMs like
e.g. propanone (CH$_3$COCH$_3$), ethylene glycol ((CH$_2$OH)$_2$) are detected in the inner protostellar
environments (e.g. Maury et al. 2014; J{\o}rgensen et al. 2016) but not yet in the shock. More observational data from hot corinos and shock regions are needed to establish whether these results are general and why.

Following Jaber et al. (2014) and Mendoza et al. (2014), we
searched for correlations between pairs of COMs detected in L1157-B1. In addition to the already known correlation
between methyl formate and dimethyl ether (Jaber et al. 2014), we found {bf good  correlations between ethanol, methanol  and glycolaldehyde} (Fig.~3).
Although these species have been detected  in only three Solar-type hot corinos, the
correlation spans two orders of magnitude and it is robust in both cases.

\begin{figure}
  \begin{center}
    \includegraphics[width=0.75\columnwidth]{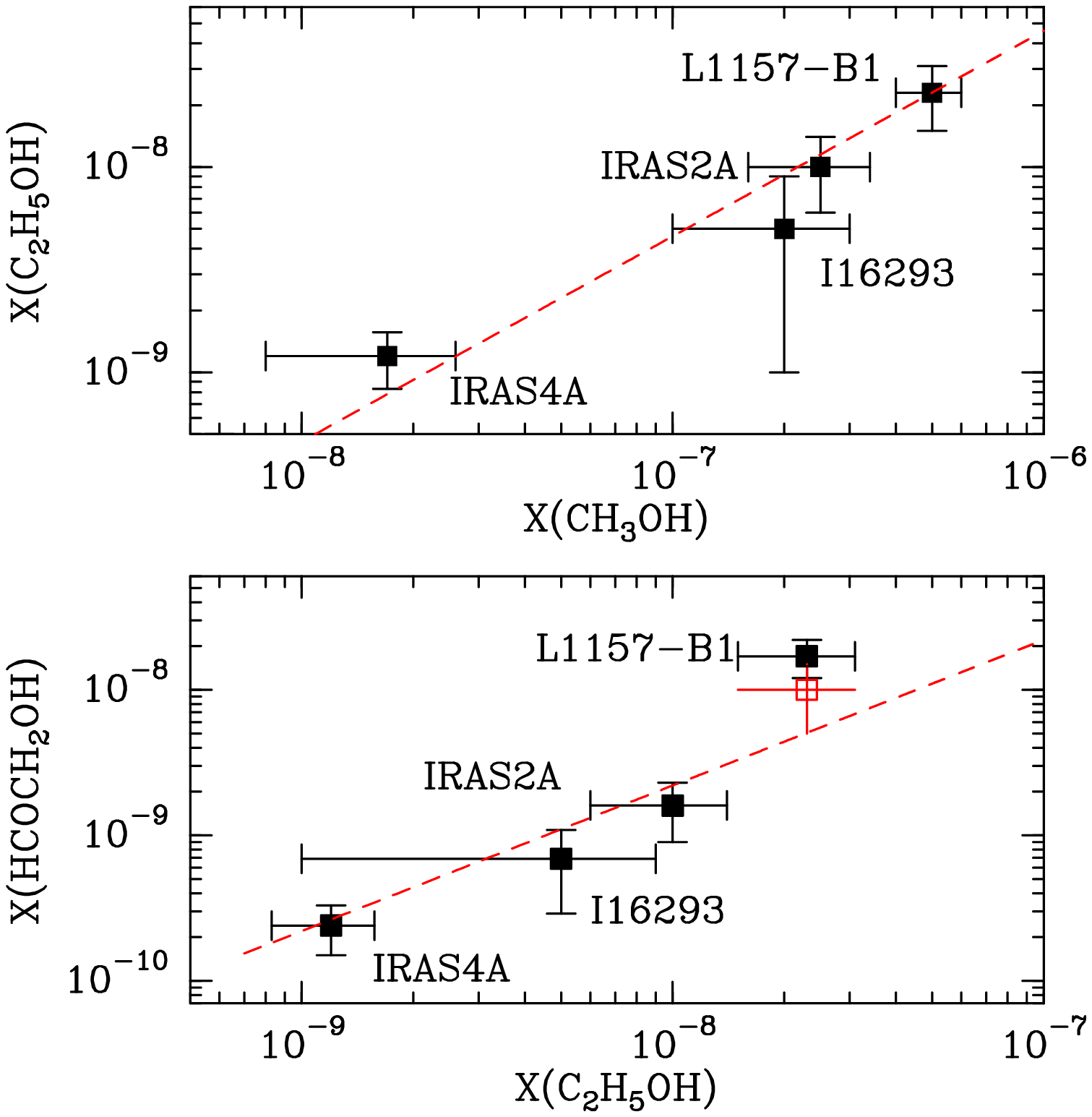}
    \caption[]{(top)~Ethanol abundance as a function of methanol abundance  in Hot Corino
      sources and L1157-B1; the relation X(C$_2$H$_5$OH)= 0.046 X(CH$_3$OH) is marked by the
      red dashed line. (bottom)~Same for glycolaldehyde and ethanol. We have plot the Glycolaldehyde abundance estimated from lines with $E_{up}\leq 55\K$ (red empty square). The relation
      X(HCOCH$_2$OH)= 0.22 X(C$_2$H$_5$OH) is marked by the red dashed line.
       }\label{fig:correlations}
  \end{center}
\end{figure}

\subsection{Formation routes for ethanol and glycolaldehyde}
The correlations methanol-ethanol and ethanol-glycolaldehyde
are intriguing and require further investigation. Indeed, in some cases,
correlations have provided a precious key to unveil the formation
route of some species. For example, Jaber at al. (2014) found that the methyl formate
and dimethyl ether abundances are very tightly correlated, with an
abundance ratio close to unity. This fact led  Balucani et al. (2015) to propose
a new formation route  methyl formate from dimethyl ether in the gas phase.
On the contrary, the tight correlation observed between formamide and isocyanic
acid (Lopez-Sepulcre et al. 2015) is still not understood.

Hydrogenation of CO on the grain surfaces is one of the few methanol formation
routes for which there is consensus (e.g. Cuppen et al. 2009; Rimola et
al. 2014). The correlations methanol-ethanol and ethanol-glycolaldehyde might then suggest that ethanol and
glycolaldehyde are formed on the grain surfaces, as suggested by some experiments. However, theoretical computations
have so far been unable to support this hypothesis. Enrique-Romero et al. (2016)
have shown that the synthesis of acetaldehyde on amorphous water ices,
predicted by the same experiments to occur thanks to the combination of
HCO and CH$_3$, is in fact impossible: such a reaction produces methane and CO,
not acetaldehyde. Conversely, simple chemical modelling in L1157-B1 suggests that acetaldehyde
may be formed in gas-phase if enough C$_2$H$_5$ is sputtered off the grain mantles in the shock
(Codella et al. 2015).

With these caveats, 
it is worth 
investigating whether ethanol and
glycolaldehyde could form in the gas phase from species present on
the grain surfaces and released in the gas phase by the shock. Both
the databases KIDA\footnote{http://kida.obs.u-bordeaux1.fr} and
UMIST\footnote{http://udfa.ajmarkwick.net} report protonated ethanol C$_2$H$_5$OH$_2^{+}$ as the
precursor of ethanol. It is synthesised in the gas by radiative association between
 H$_3$O$^+$ and C$_2$H$_4$, and H$_2$O and C$_2$H$_5^+$. Since
in the shocked gas of L1157-B1 water is abundant (e.g. Lefloch et
al. 2010; Busquet et al. 2014) and the ionisation rate is relatively
high $\zeta \simeq 3\times 10^{-16}$ (Podio et al. 2015), ethanol as a gas-phase product
is therefore a viable possibility.

Chemical models by Woods et al. (2012, 2013) do explore possible gas-phase routes for the formation of glycolaldehyde. In particular, Woods et al. (2012) include in their models the gas-phase route proposed in Halfen et al. (2006). 
The authors conclude that the predicted routes are inefficient, due to the lack of available reactants in the gas phase (e.g. formaldehyde).  As a consequence of the efficient sputtering of molecular species (e.g. H$_2$CO and CH$_3$OH) from grain mantles and the warm temperatures reached, the conditions in the shocked gas of L1157-B1 are more favourable to glycolaldehyde formation. It would be worth exploring the efficiency of these gas phase routes under such conditions.

\section{Conclusion}
We have led a systematic search for COMs in the protostellar shock region L1157-B1 as part of the ASAI Large Program. The unprecedented sensitivity of the spectral survey  has allowed us to detect nine molecular species, among which H$_2$CCO, CH$_3$OCH$_3$ and HCOCH$_2$OH for the first time in a protostellar shock. Also, we confirm the presence of HCOOH, HCOOCH$_3$ and C$_2$H$_5$OH, whose previous tentative detections were based on very few lines. Excitation temperatures and molecular abundances were obtained from a rotational diagram analysis.  COMs  appear to be more abundant in the shock L1157-B1 than in typical solar-type hot corinos, by a factor of 2--10, depending on the species. Comparison of COMs abundances in L1157-B1 and hot corinos shows new correlations between ethanol and methanol, and ethanol and glycolaldehyde. Such correlations could be related to the formation route of these molecules, which remain to be established.

\section*{acknowledgements}
IRAM is supported by INSU/CNRS (France), MPG (Germany) and IGN (Spain).  This work  was supported by the CNRS program "Physique et Chimie du Milieu Interstellaire" (PCMI) and by a grant from LabeX Osug@2020 (Investissements d'avenir - ANR10LABX56).
C. Favre acknowledges funding from the French CNES agency.

\end{document}